\documentstyle[11pt]{article}
 \def\TH{\Theta}

\def\g{\gamma}  
\def\a{\alpha}
\def\b{\beta}
\def\d{\delta} 
\def\e{\epsilon}

 \def\L{\Lambda}
\def\m{\mu}
\def\n{\nu}

\def\s{\sigma}

\def\mn{{\mu\nu}}
\def\ab{{\alpha\beta}}
\def\be{\begin{equation}}
\def\ee{\end{equation}}

\def\bea{\begin{eqnarray}}
\def\eea{\end{eqnarray}}

\begin{document}

\begin{flushright} BRX TH-550 \end{flushright}

\vspace*{.3in}

\begin{center}
{\Large\bf Duality invariance of all free bosonic and\\
\vspace{.1in} fermionic gauge fields}

{\large S.\ Deser$^1$ and D.\ Seminara$^2$}

{\it $^1$Department of Physics\\ Brandeis University\\
Waltham, MA 02454, USA\\{\tt deser@brandeis.edu}

$^2$Dipartimento di Fisica, Polo Scientifico,\\ Universit\`{a} di
Firenze;\\ INFN Sezione di Firenze Via G. Sansone 1, \\50019 Sesto
Fiorentino, Italy\\ {\tt seminara@fi.infn.it}}

\end{center}

\begin{abstract}
We give a simple general extension to all free bosonic and
fermionic massless gauge fields of a recent proof that spin 2 is
duality invariant in flat space.  We also discuss its validity in
(A)dS backgrounds and the relevance of supersymmetry.
\end{abstract}


Recently \cite{henneaux}, it was shown that the duality invariance
originally established \cite{teitelboim} for Maxwell theory in
$D$=4, extends to free massless spin 2.  It was also conjectured
there that higher spin gauge bosons, and perhaps General
Relativity (GR) might enjoy a similar property. In this Note, a
simple method is used to exhibit the duality invariance of all
free gauge fields, bosonic and fermionic; some implications of
supersymmetry and the extent of flat space duality in (A)dS are
also discussed.

We begin with bosons. The key to our derivation is the remark that
since free gauge field actions are (abelian) gauge invariant, they
can be uniformly expressed--after elimination of constraints--in
terms of the fundamental spatial gauge invariant symmetric
transverse-traceless (TT) conjugate variables \cite{waldron}
 $(\pi^{ij\cdots}_{TT}\, , \ \ q^{TT}_{ij\cdots})$,
\begin{equation}
\label{formula1}
\partial_i \pi^{ij\cdots}_{TT}\equiv \partial_i
q^{TT ~i}_{\ \ \ \ \ j\cdots}\equiv 0 \; ; \ \ \ \   \pi^{\ \ \ \
\ \ i\cdots}_{TT\ i} \equiv q^{TT ~i}_{\ \ \ \ \ ~i\cdots} \equiv
0 \; .
\end{equation}
 In this connection, we recall that only the above dynamical variables
can be meaningfully varied; neither constraint, gauge nor Lagrange
multipliers play any role here.  Only spatial indices appear ($s$
of them for spin $s$) and we drop the ``TT" henceforth.  The
obvious extension of the Maxwell action is
$$
\label{formula1a} {I}_s=\int d^4 x \left[
\pi^{ij\cdots}~\dot{q}_{ij\cdots}- H(\pi ,q)\right] \eqno{(2{\rm
a})}
$$
$$
\label{formula2b} H=\frac{1}{2}
\left[\pi^{ij\cdots}~~\pi_{ij\cdots}+
B_{ij\cdots}~~B^{ij\cdots}\right] \equiv \frac{1}{2} \;(\pi^2 +
B^2) \; , \eqno{(2{\rm b})}
$$
$$
\label{formula2c}
B_{ij\cdots}=\frac{1}{2s}\left(\epsilon_{i}^{~~lk}
\partial_l q_{kj\cdots}+
\epsilon_j^{~~l k}\partial_l q_{i k\cdots}+\cdots\right) \equiv
(\Theta q)_{ij\cdots} \eqno{(2{\rm c})}
$$
Note the extended curl operation\footnote{Actually, the integral
of $B^2$ is--as in Maxwell--equivalent to that of $(\nabla q)^2$,
but we retain the historical ``magnetic" $(\nabla\times q)^2$
notation, which preserves the index ranks of the fields in (and
only in) $D$=4 and points directly to duality.} on each index,
suitably symmetrized and normalized. Clearly, in order to preserve
the Hamiltonian, the desired transformation must be a rotation:
\setcounter{equation}{2}
\begin{equation}
\label{formula3} \delta \pi  =B ,\ \ \ \ \ \ \ \delta B =-\pi \; .
\end{equation}
 But one must first show that $\d B $ is indeed a transformation that
can be implemented by some $\d q $ and that the resulting change
in $(\pi , q)$  is canonical, namely symplectic $(\int \pi
\dot{q})$ form-preserving. As in Maxwell, this is exhibited by
\begin{equation}
\label{formula4} \delta q =-\nabla^{-2} (\Theta \pi )\ \ \ \
\Rightarrow\ \ \ \ \delta B =-\pi
\end{equation}
where $\TH$ is the generalized curl of (2c).  It is easy to check
that (3,4) is indeed canonical:  because the operator
$\nabla^{-2}\Theta$ is hermitian, the integrands in $\int
B\dot{q}$ and  $\int \pi \nabla^{-2} (\Theta \dot{\pi})$ are total
time derivatives. This completes our proof of flat space free
field bosonic duality for all spin $\geq 1$.

The fermionic case is quite analogous.  The lowest gauge field is
the spin 3/2 Rarita--Schwinger $\psi_\m$; fortunately its duality
invariance properties were derived in \cite{deser}, to which we
refer for details.  Briefly, the basic gauge-invariant variable is
now the transverse, gamma-traceless spatial vector-spinor
$\psi_i^{Tt}$
\begin{equation}
\label{formula5}
\partial^i \psi^{Tt}_i =0\; , \ \ \  \gamma^i\psi_i^{Tt}=0 \; ,
\ \ \ \gamma^i \gamma^j \psi^{Tt}_{ij} \equiv \psi^{Tt}_{ii} = 0
\; .
\end{equation}
 The ``field strength" is
\begin{equation}
\label{formula6} f_{\mu\nu} \equiv \partial_\mu
\psi_\nu-\partial_\nu\psi_\mu \; ,
\end{equation}
which obeys the (first order) field equations
\begin{equation}
\label{formula7} f_{\mu\nu}+\gamma_5 \tilde{f}_{\mu\nu}=0 \; , \ \
\ \ \ \gamma^2_5 = - 1 \; .
\end{equation}
 The theory is invariant under two separate transformations:
``bosonic" duality with respect to the world index $(f_\mn
\rightarrow \tilde{f}_\mn \equiv \frac{1}{2}\: \e_\mn~\!\!^{\ab}
f_{\a \b } )$ and ``fermionic" -- chiral-invariance under
$\g_5$-transformations on the spinor index.  The spin $(3/2 + n)$
extension scarcely needs elaboration, since it merely increases
the number of bosonic, ``Tt" indices, $\psi_i \rightarrow \psi_{ij
\ldots}$, and their tracelessness is guaranteed by (5).  Since the
gauge invariant action is again of ``bosonic" form (apart from
being first order), world index duality follows as for bosons.

We conclude our free field considerations with brief remarks on
two topics: supersymmetry and duality in constant curvature
(rather than merely flat) backgrounds. Supersymmetry (SUSY) links
the duality invariances\footnote{Not all invariances ``propagate"
like this, conformal invariance being the simplest counterexample:
only some ``tuned" higher spins retain it in constant curvature
spaces \cite{deserwaldron}. The generic criterion is commutation
of the invariance with the supercharges.} of adjoining $(s,
s+1/2)$ systems. Hence the ladder that starts at Maxwell (or
Rarita--Schwinger) is extended to all higher rungs by SUSY; this
is hardly surprising since all massless spins in $D$=4 have
exactly 2 helicities and duality just expresses their mutual
rotation.  Can free fields duality be extended to constant
curvature, (A)dS, spaces?  That of Maxwell is obvious, since it is
conformally invariant, and (A)dS is conformally flat. For $s=3/2$,
duality invariance in AdS $(\L < 0)$ was already proven in
\cite{deser}. There, the Rarita--Schwinger action necessarily
acquires a ``mass" term $\sim \sqrt{-\L} \; \psi_\m \s^\mn
\psi_\n$, in order to retain gauge invariance \cite{townsend}. The
field strength is now defined as the covariant curl
\begin{equation}
\label{formula8} f_{\mu\nu}={\mathcal{D}}_\mu
\psi_\nu-{\mathcal{D}}_\nu\psi_\mu \; , \ \ \  \ {\mathcal{D}}_\mu
\equiv D_\mu + \frac{1}{2} \sqrt{-\Lambda} \: \gamma_\mu \; , \ \
\ \  [{\cal D}_\m , {\cal D}_\n ] = 0 \; .
\end{equation}
 Commutation of the ${\cal D}_\m$ essentially reduces the process to
the flat space one and duality (but not chirality) invariance is
maintained.  Since supersymmetry is still present, for example in
linearized cosmological supergravity, the appropriate bosonic
linearization of the latter, (massless) $s=2$ in AdS, should also
be duality invariant. However, this result may not extend to
higher spins (some recent reviews of higher spins in (A)dS are
given in \cite{bouatta}), where (8) may not be applicable.

In summary, we have extended to all spins the duality invariance
established in \cite{henneaux,teitelboim,deser} for gauge fields
of spins (2,1,3/2) respectively. Our main simplification has been
to formulate the gauge invariant actions solely in terms of the
two time local ``TT" bosonic, and corresponding fermionic,
variables. The form of the field variations is then uniformly seen
to keep the $\frac{1}{2} \int (\pi^2 + {\bf B}^2)$ Hamiltonians $(
\bf{B} = \mbox{\boldmath$\nabla$} \times \bf{q})$
rotation-invariant, without affecting the symplectic $\int
\pi\dot{q}$ form.  We then used SUSY arguments to back up these
results and to extend them to AdS backgrounds, where possible. Our
results may also be of interest for higher spin fields, where a
vast literature already exists on related topics \cite{boulanger}.

We have not touched here on the important question of whether GR
shares (deformed) duality invariance with its linearized spin 2
limit, a problem we hope to discuss elsewhere.

We thank A.\ Waldron for useful discussions. This work was
supported in part by NSF grant PHY04-01667.

\end{document}